\documentclass[twocolumn,aps,nofootinbib]{revtex4}

\usepackage{graphicx}
\usepackage{epstopdf}
\usepackage{latexsym}
\usepackage{amssymb}
\usepackage{color}
\usepackage{mathrsfs}

\usepackage[center]{subfigure}

\begin{document}

 \newcommand{\bq}{\begin{equation}}
 \newcommand{\eq}{\end{equation}}
 \newcommand{\bqn}{\begin{eqnarray}}
 \newcommand{\eqn}{\end{eqnarray}}
 \newcommand{\nb}{\nonumber}
 \newcommand{\lb}{\label}
\newcommand{\PRL}{Phys. Rev. Lett.}
\newcommand{\PL}{Phys. Lett.}
\newcommand{\PR}{Phys. Rev.}
\newcommand{\CQG}{Class. Quantum Grav.}

\title{No static regular black holes in Einstein-complex-scalar-Gauss-Bonnet gravity}

\author{Kai Lin $^{1, 2}$}
\email{lk314159@hotmail.com}

\author{Shaojun Zhang$^{3, 4}$}
\email{sjzhang84@hotmail.com}

\author{Chao Zhang$^{5}$}
\email{C$\_$Zhang@baylor.edu}

\author{Xiang Zhao$^{5}$}
\email{Xiang$\_$Zhao@baylor.edu}

\author{Bin Wang$^{6, 7}$}
\email{wang$\_$b@sjtu.edu.cn}

\author{Anzhong Wang$^{5}$\footnote{ Corresponding Author}}
\email{Anzhong$\_$Wang@baylor.edu}

\affiliation {${1}$  Hubei Subsurface Multi-scale Imaging Key Laboratory, Institute of Geophysics and Geomatics, China University of Geosciences, Wuhan 430074, Hubei, China}
\affiliation{${2}$ Instituto de F\'isica, Universidade de S\~ao Paulo, S\~ao Paulo, Brazil}
\affiliation{$^3$ Institute for Advanced Physics $\&$ Mathematics, Zhejiang University of Technology, Hangzhou 310032, China}
\affiliation{$^4$ United Center for Gravitational Wave Physics,  Zhejiang University of Technology, Hangzhou 310032, China}
\affiliation{$^5$ GCAP-CASPER, Physics Department, Baylor University, Waco, TX 76798-7316, USA}
\affiliation{$^6$ Center for Gravitation and Cosmology, Yangzhou University, Yangzhou 225009, China}
\affiliation{$^{7}$ School of Physics and Astronomy, Shanghai Jiao Tong University, Shanghai 200240, China}

\date{\today}

\begin{abstract}
In this brief report, we investigate the existence of 4-dimensional static spherically symmetric black holes (BHs) in the Einstein-complex-scalar-Gauss-Bonnet (EcsGB) 
gravity with an arbitrary potential $V(\phi)$ and a coupling $f(\phi)$ between the scalar field $\phi$ and the Gauss-Bonnet (GB) term. We find that static regular BH 
solutions with  complex scalar hairs do not exist. This conclusion does not depend on  the coupling between the GB term and the scalar field, nor on the scalar potential 
$V(\phi)$ and the presence of a cosmological constant $\Lambda$ (which can be either positive or negative), as longer as the scalar field remains complex and is regular  
across the horizon.  
 
\end{abstract}


\maketitle

\section{Introduction}
\renewcommand{\theequation}{1.\arabic{equation}} \setcounter{equation}{0}

Although Einstein's general relativity (GR) has achieved a great success in physics and astronomy, especially after the recent detections of gravitational waves \cite{Ref1,GWs,GWs19a,GWs19b,LIGO} and the shadow of the M87 black hole
(BH)  \cite{EHT},  the theory is still facing some challenges, such as those related to dark energy, dark matter and quantum gravity. Therefore, various modified gravitational theories have been proposed, including  $f(R)$ gravity \cite{fR1,fR2}, Einstein-{\AE}ther ($\ae$-) theory \cite{AE1,AE2,Oost18,Foster07,Yagi14,Lin19,Zhao19,Zhang19}, Ho\v{r}ava-Lifshitz gravity \cite{HL1,HL2,HL3,Wang17}, scalar-tensor gravity \cite{ST1,ST2,ST3,ST4a,ST4b,ST4c,ST4d}, and so on. $f(R)$ gravity generalizes  the Einstein-Hilbert action to an arbitrary function of the Ricci scalar. This generalization may be able to explain the accelerated expansion and structure formation of the universe without adding unknown forms of dark energy or dark matter. Einstein-{\AE}ther gravity introduces a timelike dynamical unit vector field coupled with metric. This vector field breaks the Lorentz symmetry of the theory. It is then possible to study the preferred frame effects. Ho\v{r}ava-Lifshitz gravity treats time and space unequally at high energy level. In  scalar-tensor gravity, a scalar field and a tensor field mediate the gravitational interaction. At present, no any modified theory of gravity can replace GR, but we could  still  be able to catch a glimpse of dawn to resolve the open problems in theoretical physics, astrophysics and cosmology by exploring them. This also provides us an  excellent opportunity  to understand GR in different angles.  

As a special scalar-tensor theory, the Einstein-scalar-Gauss-Bonnet  (EsGB) gravity has attracted lots of attention recently. One of the main motivations is that, as we just enter  the era to explore the strong field regime of gravity through the direct detections of gravitational waves and BHs, understanding the effects of higher-order curvature terms becomes more urgent and important. However, the inclusion of such terms often leads to the well-known ghost problem \cite{Stelle}. The existence of ghosts is closely  related to the fact that  higher-order curvature terms contain time-derivatives higher than two. In the quadratic case, for example,  the field equations  are  generally fourth-orders. Then,  following the powerful theorem due to Mikhail Vasilevich Ostrogradsky,  who established it   in 1850  \cite{Ostrogradsky}, such a system is generically not stable, unless the degenerate conditions are satisfied  \cite{Woodard15,Wang17}.  The Gauss-Bonnet (GB) term belongs precisely to the latter:Due to a particular combination of the Riemann and Ricci tensors, it  contains no more than the second-order derivative terms, although it consists of the quadratic terms of the Riemann tensor.  As an immediate result,  
 the theory is ghost-free.   However, when it is minimally coupled to the Einstein-Hilbert action, it becomes a topological term in 4-dimensional spacetimes, and has no contributions to the field equations. To make this term meaningful 
 in  4-dimensional spacetimes, one way is to consider it coupling non-minimally with a scalar field, partially motivated from string/M-Theory \cite{string}. 
 
Along this vein, EsGB gravity has been extensively studied in the past decade or so \cite{Berti15,HR15,Macedo20}. In particular,    both static \cite{realscalar1,realscalar2,realscalar3} and stationary \cite{realscalar4} BH  
solutions have been found, through the so-called spontaneous scalarization, by which  an initial GR BH spontaneously grows hairs through a tachyonic instability.  The circulation of the well-known no-hair theorems \cite{RW71,Bek72,Teit72,Bek95} is through the non-minimally coupling between the scalar and the GB term, which effectively produces a negative mass-squared term in the scalar field perturbations.  It should be noted that 
spontaneous scalarization has been known for a long time in the studies of neutron stars in a particular class of scalar-tensor theories \cite{DEF93}, in which the effective mass was provided by matter terms of the theory. However,
in the EsGB theory, it is provided by the coupling of the GB curvature term with a scalar field. 

It should be also noted that in the above studies of the BH spontaneous scalarization, the scalar field was always assumed to be real. In this short Note, we generalize these studies to a complex scalar field, 
and then ask ourselves if BHs can still exist. The answer is somehow a bit surprising and negative: {\em No static regular BHs with positive temperatures exist in the Einstein-complex-scalar-Gauss-Bonnet  (EcsGB) 
theory with an arbitrary coupling  $f(\phi)$ between  the GB term and the scalar field and an  arbitrary potential $V(\phi)$, as longer as the scalar field remains complex and regular across the horizon}. Since $V(\phi)$ is arbitrary, this also includes the case with a non-vanishing cosmological constant $\Lambda$, which  can be either positive or negative.

Before proving our above claim, we note that  {in the Standard Model (SM) of particle physics, the  Higgs field is a complex scalar with the $SU(2)_L$ symmetry \cite{Higgs64,EB64,GHK64}, 
which is the only scalar boson  that has been proved so far to exist   in our Universe  \cite{LHC12a,LHC12b}. 
Applying it to cosmology, it was shown that the Higgs field with a non-minimally coupling can serve as the inflaton field \cite{BS08,BD08}, and is consistent with all the observations 
 \cite{Planck2018_Inflation}. It should be also noted that the Higgs field  minimally coupled to gravity  usually produces   matter
fluctuations that are many orders of magnitude larger than those observed, unless a fine-tuned value of the Higgs self-coupling constant $\lambda$ is invoked \cite{AL83,JR19}.}

In addition, boson stars in the framework of EcsGB theory were studied recently in \cite{BM17,BD19}, while boson stars in GR coupled with a (complex) scalar field but without the GB term
have been extensively studied (see, for example, \cite{Jetzer92,LP17,Li20} and references therein), since the pioneer work of Kaup \cite{Kaup68}, and Ruffini and Bonazzola \cite{RB69}. 
  {Among these works,  in the studies
of  spherically symmetric static configurations, without loss of the generality (as to be shown in the next section),  the majority 
adopted the  standard harmonic ansatz  of the   complex scalar field,   $\phi(t, r) = e^{iw t}\Psi(r)$ \cite{Lee87,FLP87}. In this paper, we shall also adopt this ansatz.}

\section{No-go theorem of Static BHs in EcsGB Theory}
\renewcommand{\theequation}{2.\arabic{equation}} \setcounter{equation}{0}

The general action of the  EcsGB theory  takes the form,  
\bq
\lb{Action}
S=S_g + S_{m}, 
\eq
where
\bqn
\lb{eq2.1}
S_g&=&\int dx^4\sqrt{-g}\left(\frac{c_1R-2\Lambda}{2\kappa}+\alpha{\cal L}_{GB}+{\cal L}_{\phi}\right),\nb\\
S_m &=& \int dx^4\sqrt{-g}{\cal L}_{m}\left(g_{\mu\nu}, \psi\right), ~~~
\eqn
with $\kappa\equiv8\pi G$, $\psi$ denoting collectively the matter fields,  and
\bqn
\lb{eq2.2}
{\cal L}_{GB}&\equiv& f(\phi\phi^*){\cal G},\nb\\
{\cal L}_\phi&\equiv&-\nabla^\mu\phi\nabla_\mu\phi^*-V(\phi\phi^*),\nb\\
{\cal G}&\equiv&R^2+R_{\mu\nu\rho\sigma}R^{\mu\nu\rho\sigma}-4R_{\mu\nu}R^{\mu\nu},
\eqn
where $\Lambda$ is the cosmological constant, $c_1$ and $\alpha$ are coupling constants, an $R_{\mu\nu\rho\sigma},\; R_{\mu\nu}$ and $R$ are the Riemann, Ricci tensors and Ricci scalar, respectively. 
$\nabla_{\mu}$ denotes the covariant derivative with respect to the metric $g_{\mu\nu}$, and $g \equiv \text{det}(g_{\mu\nu})$. It's worth noting that the action requires $c_1\ge0$ to avoid $\phi$ being a phantom field. For the $c_1>0$ case, we can directly redefine the parameters so that $c_1=1$. On the other hand, the GR terms will vanish when $c_1=0$, while the effects of the scalar field {vanish}  and the action {reduces} to GR when $c_1\to\infty$.
Here $V$ and $f$ are functions of $\phi\phi^*=|\phi|^2$. We will adopt the notations $V^{(1)}\equiv\partial V/\partial \left(|\phi|^2\right)$, $f^{(1)}\equiv\partial f/\partial \left(|\phi|^2\right)$ and $f^{(2)}\equiv\partial^2 f/{\partial \left(|\phi|^2\right)}^2$. 
For the massive scalar field without self-interaction, we have   $V=m^2\left|\phi\right|^2$, where $m$ denotes the mass of the scalar field. In this paper, we shall consider the general case where $V$ is an arbitrary function of $|\phi|^2$.  
Similarly, we also do not impose any constraint on the  form of  $f(\phi)$, but consider it as an arbitrary function of $\left|\phi\right|^2$. 
 
With the above in mind, we find that the scalar field equations are given by,  
\bqn
\lb{ScalarField}
\Box \phi= \left(V^{(1)} - \alpha{\cal G}f^{(1)}\right)\phi,
\eqn
where $\Box \equiv g^{\mu\nu}\nabla_{\mu}\nabla_{\nu}$, while the gravitational field equations take the forms,   
\bqn
\lb{GravitationalField}
{ c_1\left(R_{\mu\nu}-\frac{1}{2}g_{\mu\nu}R \right) + \Lambda g_{\mu\nu}}&=&\kappa\Big(\alpha T^{GB}_{\mu\nu}+T^\phi_{\mu\nu}\nb\\
&&  ~~~~ + T^m_{\mu\nu}\Big),
\eqn
where $T^m_{\mu\nu}$ denotes the energy-momentum tensor of the matter fields, and
\bqn
\lb{GravitationalFieldTensor}
T^{GB}_{\mu\nu}&=&4(\nabla_\mu\nabla_\nu f)R-4g_{\mu\nu}(\nabla_\rho\nabla^\rho f)R\nb\\
&&-8(\nabla^\rho\nabla_\nu f)R_{\mu\rho}-8(\nabla^\rho\nabla_\mu f)R_{\nu\rho}\nb\\
&&+8(\nabla^\rho\nabla_\rho f)R_{\mu\nu}+8g_{\mu\nu}(\nabla^\rho\nabla^\sigma f)R_{\rho\sigma}\nb\\
&&-8(\nabla^\rho\nabla^\sigma f)R_{\mu\rho\nu\sigma}.\nb\\
T^\phi_{\mu\nu}&=&\nabla_\mu\phi\nabla_\nu\phi^*+\nabla_\nu\phi\nabla_\mu\phi^*-g_{\mu\nu}V(|\phi|^2)\nb\\
&&-g_{\mu\nu}\nabla^\rho\phi\nabla_\rho\phi^*.
\eqn

In this paper, we shall consider the vacuum case, $T^m_{\mu\nu} = 0$, and static spacetimes with spherical symmetry, for which the metric can be cast in the general form, 
\bqn
\lb{Metric}
ds^2=-e^{{A(r)}}dt^2+e^{{B(r)}}dr^2 +r^2\left(d\theta^2+\sin^2\theta d\varphi^2\right). ~~
\eqn
In such spacetimes, let us consider the solution $\phi(t, r) = \Phi(t)\Psi(r)$, for which Eq.(\ref{ScalarField}) yields, 
\bqn
\lb{eq2.4}
&& \frac{e^A\Psi''}{e^B\Psi}+\left(4+rA'-rB'\right)\frac{e^A\Psi'}{2re^B\Psi}+\left\{-e^AV^{(1)}\right.\nb\\
&& ~~~~~~ -e^{A-B}\frac{2\alpha f^{(1)}}{r^2e^B}\left(2A''+A'^2\right)\left(e^B-1\right)\nb\\
&& ~~~~~~ \left.+e^{A-B}\frac{2\alpha f^{(1)}}{r^2e^B}\left(e^B-3\right)A'B'\right\} \nb\\
&&=\frac{\ddot\Phi(t)}{\Phi(t)} \equiv  -w^2.
\eqn
Clearly, for the above equation to be consistent,  $w$ must be a constant, and the general solution of $\Phi$ is given by,
$\Phi(t) = \Phi_0e^{-iw (t + t_0)}$, where $\Phi_0$ and $t_0$ are the two integration constants. To have the spacetimes stable, we must assume that $\omega$ is real and non-zero. 
On the other hand, redefining $t' = t + t_0$, without loss of the generality, we can always  set $t_0 = 0$.  Meanwhile,  we can absorb the constant $\Phi_0$ to $\Psi(r)$
 \footnote{In the framework of GR coupled with a scalar field,
it was found that stationary scalar clouds around a rotating BH can exist \cite{Hod12}, while in the spherically symmetric case, such a claud exists around a Schwarzschild BH 
only in time-dependent case \cite{Barranco11},  in which  the scalar field decays slowly for a long time. If the mass of the scalar field is ultralight,  the decay time can be comparable with the age of our Universe.
As a result, such  ultralight scalar fields can be considered as possible candidates for  dark matter \cite{HBG00,NSS01,Briscese11,Barranco12,HOTW17}. For more details, see the review articles \cite{SRM14,Marsh16}.
However, in this paper, we are mainly interested in static black hole solutions, so here we shall not consider this case.}, so finally we have $\phi=\Psi(r)e^{-iw t}$.
  {Without loss of the generality, we can further assume that $\Psi(r)$ is real \cite{Lee87,FLP87}. }  Hence, we obtain   the standard harmonic ansatz  of the   complex scalar field  \cite{Jetzer92,LP17,BM17,BD19}. 
Thus, Eq.(\ref{eq2.4}) becomes 
\bqn
\lb{FieldEquationA}
&& \Psi''+\left(4+rA'-rB'\right)\frac{\Psi'}{2r}+\left\{\frac{w^2-e^AV^{(1)}}{e^{A-B}}\right.\nb\\
&& ~~~~~~ -\frac{2\alpha f^{(1)}}{r^2e^B}\left(2A''+A'^2\right)\left(e^B-1\right)\nb\\
&& ~~~~~~ \left.+\frac{2\alpha f^{(1)}}{r^2e^B}\left(e^B-3\right)A'B'\right\}\Psi = 0, 
\eqn
while 
  the field equations  (\ref{GravitationalField}) have only two independent components, given, respectively,  by, 
 \bqn
\lb{FieldEquationB}
&& \left[c_1e^Br+8\alpha\kappa(e^B-3)\Psi\Psi' {f^{(1)}}\right]e^AB'\nb\\
&& ~~~~~~ -16\alpha\kappa e^A\Psi\Psi''\left(e^B-1\right)f^{(1)}-\kappa e^{A+2B}Vr^2\nb\\
&& ~~~~~~ -\kappa \left[w^2e^{2B}r^2+32\alpha e^A\left(e^B-1\right)\Psi'^2f^{(2)}\right]\Psi^2\nb\\
&& ~~~~~~ +e^{A+B}\left[c_1\left(e^B-1\right)-\Lambda r^2e^B\right]\nb\\
&& ~~~~~~ -\kappa e^A\left[e^Br^2+16\alpha\left(e^B-1\right)f^{(1)}\right]\Psi'^2 = 0, ~~~~~~~~\\
\lb{FieldEquationC}
&& \left[c_1r+8\alpha\kappa (1-3{e^{-B}})\Psi\Psi'f^{(1)}\right]e^AA'\nb\\
&&~~~~~~ +\kappa e^Br^2\left(e^A {V}-w^2\Psi^2\right)\nb\\
&&~~~~~~ +e^A\left[c_1+e^B(\Lambda r^2-c_1)-\kappa r^2\Psi'^2\right]=0.
\eqn

 Let us first note that when the scalar field is real ($w = 0$), the Schwarzschild-de Sitter solution is indeed a solution of Eqs.(\ref{FieldEquationA})-(\ref{FieldEquationC}), if we further set $\Psi = \Psi_0$
and assume that the scalar field stays at its minimum, i.e., $V^{(1)}(\Psi_0) = 0 = f^{(1)}(\Psi_0)$, where $\Psi_0$ is a real constant.

 {However, as longer as $w \not= 0$, the situation will be completely different. 
To show this,}  we first assume that a regular BH exists, and is located at $r = r_h$,  so that $e^{A(r_h)}=e^{-B(r_h)}=0$. Additionally, we assume that the temperature of BH is positive and finite, 
and that  the scalar field   is also regular across the horizon.  Then, near the horizon the metric and the scalar field can be expanded as \cite{realscalar1,Macedo20}, 
\bqn
\lb{Expand}
e^A&=&\sum_{i=1}^{\infty}F_{i}\left(r-r_h\right)^i,\nb\\
e^{-B}&=&\sum_{i=1}^{\infty}H_{i}\left(r-r_h\right)^i,\nb\\
\Psi&=&\sum_{i=0}^{\infty}\Psi_{i}\left(r-r_h\right)^i,
\eqn
where $F_i$, $H_i$ and $\Psi_i$ are constant {coefficients.} Substituting the above functions into the field equations, and then expanding them near the horizon $r_h$, we find
\bqn
\lb{Series}
w^2\Psi_0r_h^2 +{\cal O}\left(r-r_h\right)&=&0,\nb\\
\kappa w^2\Psi_0^2r_h^2 +{\cal O}\left(r-r_h\right)&=&0,\nb\\
 \kappa w^2\Psi_0^2r_h^2 +{\cal O}\left(r-r_h\right)&=&0. ~~~~
\eqn
Thus, to the zeroth-order of $r - r_h$, we must have $w\Psi_0 r_h=0$.
 Therefore, there are three possibilities: $r_h=0, \; \forall \; w, \; \Psi$;  $w=0, \; \forall \; \Psi, \; r_h$; and $\Psi_0= 0, \;  \forall \; w, \; r_h$. When $r_h=0$, there are no BH solutions. 
 For the $w=0$ case, the scalar field becomes real, and in this case both static and rotating  BHs  exist  \cite{realscalar1,realscalar2,realscalar3,realscalar4}. 
 For BHs with complex scalar hairs, the above shows that we must have  $\Psi_0\equiv\Psi(r=r_h)=0$.

After substituting $\Psi_0=0$ into  Eqs.(\ref{FieldEquationA})-(\ref{FieldEquationC}), to the the first-order of $r - r_h$,   we find,  
\bqn
\lb{Series2}
 F_1&=& \frac{c_1-(\kappa V(0)+\Lambda) r_h^2}{c_1r_h}, \nb\\ 
 H_1&=& -\frac{c_1r_hw^2}{c_1-(\kappa V(0)+\Lambda) r_h^2}, ~~~~
 \eqn
from which we obtain  $F_1H_1=-w^2<0$. Then, the temperatures of such BHs $T_h=\sqrt{F'(r_h)H'(r_h)}/4\pi$ 
become   imaginary, which is physically unacceptable.  In other words, 
static spherically symmetric BHs with a complex scalar field does not exist in the EcsGB theory.

\section{Concluding Remarks}

In this short brief report, we have shown that static regular BHs in EcsGB theory do not exist, as longer as the scalar field remains complex and the BH is regular and has a positive temperature.
This conclusion in particular is independent of  the coupling $f(|\phi|^2)$ between the GB term ${\cal L}_{GB}$ [defined by Eq.(\ref{eq2.2})],  and the complex scalar field $\phi$.  It is also 
independent of the potential $V(|\phi|^2)$ of the complex scalar field,  and the presence of  a non-vanishing cosmological constant $\Lambda$, which  can be either positive or negative.  
Therefore, the no-go theorem is quite general. 

To circulate the theorem, one way is to consider time-dependent spherically symmetric BHs, similar to the case with a real scalar field studied in  \cite{HBG00,NSS01,Briscese11,Barranco11,Barranco12,HOTW17}.
Another extension is to consider BHs with rotation. Again, in the real scalar field case, such stationary rotating BHs exist \cite{Hod12}. In addition, when the scalar field is complex, but without its coupling to the GB term, it was also found that    {rotating} BHs with complex scalar hairs exist \cite{HR14}. Therefore, it would be very interesting to show that    {rotating} BHs in the EcsGB theory also exist. We hope to report our findings in these  directions soon.

\begin{acknowledgments}

This work is partially supported by 
 the National Natural Science Foundation of China (NNSFC) under Grant Nos. 11675145, 11805166, and  11975203.

\end{acknowledgments}


\begin{thebibliography}{nbound}



\bibitem{Ref1} B.P. Abbott, {et al.,} [LIGO/Virgo Scientific Collaborations], Observation of Gravitational Waves from a Binary Black Hole Merger, Phys. Rev. Lett. {\bf 116},  061102 (2016).



  
\bibitem{GWs}  B.P. Abbott, {et al.,} [LIGO/Virgo Collaborations], GWTC-1: A Gravitational-Wave Transient Catalog of Compact Binary Mergers Observed by LIGO and Virgo during the First and Second Observing Runs, Phys.\ Rev.\ X {\bf 9},  031040 (2019).

\bibitem{GWs19a}  B.P. Abbott, {et al.,} [LIGO/Virgo Collaborations], Open data from the first and second observing
runs of Advanced LIGO and Advanced Virgo, arXiv:1912.11716.

\bibitem{GWs19b} B.P. Abbott, {et al.,} [LIGO/Virgo Collaborations], GW190425: Observation of a Compact Binary Coalescence with Total Mass $\sim 3.4M_{\bigodot}$,
arXiv:2001.01761.

\bibitem{LIGO} https://www.ligo.caltech.edu/

\bibitem{EHT} K. Akiyama,  {et al.,} [The Event Horizon Telescope Collaboration], First M87 Event Horizon Telescope Results. I. The Shadow of the Supermassive Black Hole, Astrophys. J. L. {\bf 875} (2019) L1. 

\bibitem{fR1} W. Hu and I. Sawicki, Models of f(R) cosmic acceleration that evade solar system tests, Phys. Rev. D{\bf 76}, 064004 (2007).
\bibitem{fR2} A.D. Felice and S. Tsujikawa, f(R) Theories, Living Rev. Relativ. {\bf 13}, 3 (2010).

\bibitem{AE1} T. Jacobson, D. Mattingly, Gravity with a dynamical preferred frame, Phys. Rev. D D{\bf 64}, 024028 (2001); Einstein-aether waves, 70, 024003 (2004).
\bibitem{AE2} T. Jacobson, Einstein-{\AE}ther gravity: a status report, arXiv:0801.1547.

  \bibitem{Oost18} J. Oost, S. Mukohyama and A. Wang, Constraints on Einstein-aether theory after GW170817, Phys. Rev. D{\bf 97}, 124023 (2018).



 \bibitem{Foster07} B.Z. Foster, Strong field effects on binary systems in Einstein-aether theory, Phys. Rev. D{\bf 76}, 084033 (2007).

 \bibitem{Yagi14}   K. Yagi,  D. Blas, E. Barausse, and N. Yunes,   Constraints on Einstein-$\ae$ther theory and Ho\v{r}ava gravity from binary pulsar observations, Phys. Rev. D{\bf 89}, 084067 (2014).
 

\bibitem{Lin19} K. Lin, X. Zhao, C. Zhang, K. Lin, T. Liu, B. Wang, S.-J. Zhang, X. Zhang, W. Zhao, T. Zhu, A. Wang, Gravitational waveforms, polarizations, response functions,
and energy losses of triple systems in Einstein-aether theory, Phys. Rev. D99, 023010 (2019).

 
\bibitem{Zhao19} X. Zhao, C. Zhang, K. Lin, T. Liu, R. Niu, B. Wang, S.-J.  Zhang, X. Zhang, W. Zhao, T. Zhu, A. Wang, 
Gravitational waveforms and radiation powers of the triple  system PSR J0337+1715 in modified theories of gravity,  Phys. Rev. D{\bf 100}, 083012 (2019).
           

\bibitem{Zhang19} C. Zhang, X. Zhao, A. Wang ,B.  Wang, K. Yagi N. Yunes,
W. Zhao, and T. Zhu, Gravitational waves from the quasicircular inspiral of compact binaries in Einstein-aether theory, Phs. FRev. D{\bf 101}, 044002 (2020).

\bibitem{HL1} P. Horava, Quantum gravity at a Lifshitz point, Phys. Rev. D{\bf 79}, 084008 (2009).

\bibitem{HL2} S. Mukohyama, Ho\v rava-Lifshitz cosmology: a review, Classical Quantum Gravity 27, 223101 (2010).

\bibitem{HL3} K. Lin, S. Mukohyama, A. Wang and T. Zhu, Post-Newtonian approximations in the Ho\v rava-Lifshitz gravity with extra U(1) symmetry, Phys. Rev. D{\bf 89}, 084022 (2014).

\bibitem{Wang17}  A. Wang, Ho\v{r}ava Gravity at a Lifshitz Point: A Progress Report, Int. J. Mod. Phys. D{\bf 26} (2017) 1730014.

\bibitem{ST1} C. M. Will, Testing scalar-tensor gravity with gravitational-wave observations of inspiralling compact binaries, Phys. Rev. D{\bf 50}, 6058 (1994).
\bibitem{ST2} G. E.-Farese, Tests of Scalar-Tensor Gravity , AIP Conference Proceedings 736, 35 (2004).

\bibitem{ST3} P. Brax, A.-C. Davis, B. Li,  and H.A. Winther, Unified description of screened modified gravity,  Phys. Rev. D{\bf 86}, 044015   (2012).

 \bibitem{ST4a} X. Zhang, J. Yu, T. Liu, W. Zhao, A. Wang, 
Testing Brans-Dicke gravity using the Einstein telescope,  Phys. Rev. D{\bf 95}, 124008 (2017).


\bibitem{ST4b} X. Zhang, W. Zhao, T. Liu, K. Lin, C. Zhang, X. Zhao, S.-J. Zhang, T. Zhu, and A. Wang, 
Constraints of general screened modified gravities from comprehensive analysis of binary pulsars, 
Astrophy. J., {\bf 874},  121 (2019).

\bibitem{ST4c} X. Zhang, W. Zhao, T. Liu, K. Lin, C. Zhang, X. Zhao, S.-J. Zhang, T. Zhu, and A. Wang,
 Angular momentum loss for eccentric compact binary in screened modified gravity, JCAP {\bf 01} (2019) 019.
 
 \bibitem{ST4d} T. Liu, X. Zhang, W. Zhao, K. Lin, C. Zhang, S.-J. Zhang, X. Zhao, T. Zhu, A. Wang,
 Waveforms of compact binary inspiral gravitational radiation in screened modified gravity, Phys. Rev. D{\bf 98}, 083023 (2018).



 \bibitem{Stelle} K. S. Stelle, {\em Renormalization of higher-derivative quantum gravity}, Phys. Rev. D{\bf 16} (1977) 953.
 
 
\bibitem{Ostrogradsky}  M. Ostrogradsky, M\'emoire sur les \'equations diff\'erentielles relatives au probl\`eme des isop\'erim\`etres, Mem. Ac. St. Petersbourg, VI{\bf 4} (1850) 385.


\bibitem{Woodard15} R. P. Woodard, {\em The Theorem of Ostrogradsky}, Scholarpedia  {\bf 10}  (2015) 32243 [arXiv:1506.02210]. 
 

\bibitem{string}
M.B. Green, J.H. Schwarz and E. Witten, {\em Superstring Theory: Vol.1 $\&$ 2}, Cambridge Monographs on Mathematical Physics (Cambridge University Press,
                     Cambridge, 1999);  J. Polchinski, {\em String Theory, Vol. 1 $\&$ 2} (Cambridge University Press, Cambridge, 2001);
                      C. V. Johson, {\em D-Branes}, Cambridge Monographs on Mathematical Physics (Cambridge University Press,  Cambridge, 2003);
                      K. Becker, M. Becker, and J.H. Schwarz, {\em String Theory and M-Theory} (Cambridge University Press, Cambridge, 2007).
                      
\bibitem{Berti15}   E. Berti {\em et al.}, Testing general relativity with present and future astrophysical observations, Class. Quant. Grav. {\bf 32} (2015) 243001.

 
\bibitem{HR15}  C.A.R. Herdeiro and E. Radu, Asymptotically flat BHs with scalar hair: A review, Int. J. Mod. Phys. D{\bf 24} (2015) 1542014.


\bibitem{Macedo20}  C.F.B. Macedo, Scalar modes, spontaneous scalarization and circular null-geodesics of
BHs in scalar-Gauss-Bonnet gravity, arXiv:2002.12719. 


\bibitem{realscalar1} G. Antoniou, A. Bakopoulos and P. Kanti, Evasion of No-Hair Theorems and Novel Black-Hole Solutions in Gauss-Bonnet Theories,  Phys. Rev. Lett., {\bf 120}, 131102 (2018).

\bibitem{realscalar2} D. D. Doneva and S. S. Yazadjiev, New Gauss-Bonnet Black Holes with Curvature-Induced Scalarization in Extended Scalar-Tensor Theories, Phys. Rev. Lett., {\bf 120},  131103 (2018).

\bibitem{realscalar3} H. O. Silva, J. Sakstein, L. Gualtieri, T. P. Sotiriou and E. Berti, Spontaneous Scalarization of Black Holes and Compact Stars from a Gauss-Bonnet Coupling, Phys. Rev. Lett., {\bf 120},  131104 (2018).


\bibitem{realscalar4} P. V. P. Cunha, C. A. R. Herdeiro and E. Radu, Spontaneously Scalarized Kerr Black Holes in Extended Scalar-Tensor–Gauss-Bonnet Gravity, Phys. Rev. Lett., {\bf 123},  011101 (2019).

\bibitem{RW71} R. Ruffini and J. A. Wheeler, Introducing the black hole,  Phys. Today {\bf 24}, 30 (1971).

\bibitem{Bek72}  J. D. Bekenstein, Transcendence of the Law of Baryon-Number Conservation in Black-Hole Physics, Phys. Rev. Lett. {\bf 28}, 452 (1972).

\bibitem{Teit72}   C. Teitelboim, Nonmeasurability of the lepton number of a black hole, Lett. Nuovo Cimento {\bf 3}, 397 (1972).

\bibitem{Bek95}  J. D. Bekenstein,     Rapid Communication
Novel ``no-scalar-hair'' theorem for black holes, Phys. Rev. D{\bf 51}, R6608 (1995).




\bibitem{DEF93} T. Damour, G.Esposito-Farese,  Nonperturbative Strong-Field Effects in Tensor-Scalar Theories of Gravitation, 
Phys. Rev. Lett., {\bf 70},  2220 (1993).


\bibitem{Higgs64} P. W. Higgs, Broken symmetries, massless particles and gauge fields, Phys. Lett. {\bf 12},  132 (1964);
 BROKEN SYMMETRIES AND THE MASSES OF GAUGE BOSONS, Phys. Rev. Lett. {\bf 13}, 508 (1964)



\bibitem{EB64} F. Englert and R. Brout, BROKEN SYMMETRY AND THE MASS OF GAUGE VECTOR MESONS, Phys. Rev. Lett. {\bf 13},  321 (1964).


\bibitem{GHK64} G.S. Guralnik, C.R. Hagen and T.W.B. Kibble, GLOBAL CONSERVATION LAWS AND MASSLESS PARTICLES, Phys. Rev. Lett. {\bf 13}, 585 (1964).


\bibitem{LHC12a} ATLAS Collaboration, G. Aad {\it et al.}, Observation of a new particle in the search for the Standard Model Higgs boson with the ATLAS detector at the LHC, Phys. Lett. B{\bf 716} (2012) 1.


\bibitem{LHC12b} CMS Collaboration, S. Chatrchyan {\it et al.}, Observation of a new boson at a mass of 125 GeV with the CMS experiment atthe LHC, Phys. Lett. B{\bf 716} (2012) 30. 



\bibitem{BS08} F. Bezrukov, M. Shaposhnikov, The Standard Model Higgs boson as the inflaton, Phys. Lett. B{\bf 659} (2008) 703. 

\bibitem{BD08} F. Bauer and D. A. Demir, Inflation with Non-Minimal Coupling: Metric versus Palatini Formulations, Phys. Lett. B{\bf 665} (2008) 222.

\bibitem{Planck2018_Inflation}  Planck collaboration, Planck 2018 results. X. Constraints on inflation, arXiv:1807.06211.

\bibitem{AL83} A. D. Linde. Chaotic Inflation, Phys. Lett. B{\bf 129}, 177 (1983).

\bibitem{JR19} J. Rubio, Higgs Inflation, Front. Astron. Space Sci. {\bf 5} (2019) 50  [arXiv:1807.02376].



\bibitem{BM17} V. Baibhav and D. Maity, Boson stars in higher-derivative gravity, Phys. Rev. D{\bf 95}, 024027 (2017).

\bibitem{BD19} Y. Brihaye and L. Ducobu, Hairy black holes, boson stars and non-minimal coupling to curvature invariants, Phys.  Lett. B{\bf 795}, (2019) 135.

\bibitem{Jetzer92} P. Jetzer, Boson stars, Phys. Reports,  {\bf 220} (1992) 163.

\bibitem{LP17}  S. L. Liebling and C. Palenzuela, Dynamical boson stars, Living Rev. Relativity {\bf 15}, 6 (2012);  {\bf 20}, 5 (2017).

\bibitem{Li20} H.-B. Li, S. Sun, T.-T. Hu, Y. Song, Y.-Q. Wang, Rotating multistate boson stars, Phys. Rev. D{\bf 101}, 044017 (2020).

\bibitem{Kaup68}  D. J. Kaup, Klein-Gordon Geon, Phys.Rev. {\bf 172}, 1331(1968).

\bibitem{RB69}  R. Ruffini and S. Bonazzola, Systems of self-gravitating particles in general relativity and the concept of an equation
of state, Phys. Rev. {\bf 187}, 1767 (1969).


\bibitem{Lee87}   T. D. Lee, Soliton stars and the critical masses of black holes, Phys. Rev. D{\bf 35}, 3637 (1987).

\bibitem{FLP87} R. Friedberg, T. D. Lee, and Y. Pang, Mini-soliton stars, Phys. Rev. D{\bf 35}, 3640 (1987).

\bibitem{Hod12} S. Hod, Stationary scalar clouds around rotating BHs, Phys. Rev. D{\bf 86}, 104026 (2012).

\bibitem{Barranco11} J. Barranco, A. Bernal, J. C. Degollado, A. Diez-Tejedor,
M. Megevand, M. Alcubierre, D. Nunez, and O. Sarbach,  Are BHs a serious threat to scalar field dark matter models?
  Phys. Rev. Lett. D{\bf 84}, 083008 (2011).


\bibitem{HBG00} W. Hu, R. Barkana, and A. Gruzinov, Fuzzy Cold Dark Matter: The Wave Properties of Ultralight Particles, Phys. Rev. Lett. {\bf 85}, 1158 (2000).

\bibitem{NSS01} U. Nucamendi, M. Salgado, and D. Sudarsky, Alternative approach to the galactic dark matter problem, Phys. Rev. D{\bf 63}, 125016 (2001).

\bibitem{Briscese11} F. Briscese, Viability of complex self-interacting scalar field as dark matter, Phys. Lett. B{\bf 696}, 315 (2011).

\bibitem{Barranco12} J. Barranco, A. Bernal, J. C. Degollado, A. Diez-Tejedor, M. Megevand, M. Alcubierre, D. Nunez, and O. Sarbach,  Schwarzschild BHs can Wear Scalar Wigs, 
              Phys. Rev. Lett. {\bf 109}, 081102 (2012).

\bibitem{HOTW17}  L. Hui, J. P. Ostriker, S. Tremaine, and E. Witten, Ultralight scalars as cosmological dark matter, Phys. Rev. D{\bf 95}, 043541 (2017).
  


\bibitem{SRM14}  A. Suarez, V. H. Robles, and T. Matos, A review on the scalar field/Bose-Einstein condensate dark matter model, Astrophys. Space Sci. Proc. 38, 107 (2014).

\bibitem{Marsh16} D. J. E. Marsh, Axion cosmology, Phys. Rep. {\bf 643}, 1 (2016).
 
 \bibitem{HR14} C.A.R. Herdeiro and E. Radu, Kerr Black Holes with Scalar Hair, Phys. Rev. Lett. {\bf 112}, 221101 (2014).
 
 
 


\end{thebibliography}
\end{document}